\documentclass[aps,prl,showpacs,twocolumn]{revtex4}   
\usepackage{amsmath}    
\usepackage{graphicx}   

\begin{document}

\title{Purcell factor for point--like dipolar emitter coupling to 2D-plasmonic waveguides}
\author{J. Barthes} 
\author{G. {Colas des Francs}}
\email{gerard.colas-des-francs@u-bourgogne.fr}
\author{A. Bouhelier}
\author{J-C. Weeber}
\author{A. Dereux}
 \affiliation{Laboratoire Interdisciplinaire Carnot de Bourgogne, UMR 5209 CNRS - Universit\'e de Bourgogne, \\
9 Av. A. Savary, BP 47 870, 21078 Dijon, FRANCE}  

\date{\today}

\newcommand{\gras}[1]{\mathbf{#1}}
\begin{abstract}
We theoretically investigate the spontaneous emission of a point--like dipolar emitter located near a two--dimensional (2D) plasmonic waveguide of arbitrary
form. We invoke an explicite link with the density of modes of the waveguide describing the electromagnetic
channels into which the emitter can couple. We obtain a closed form expression for the coupling to propagative plasmon, extending thus the Purcell factor to plasmonic configurations. Radiative and non-radiative contributions to the spontaneous emission are also
discussed in details. 
\end{abstract}


\maketitle

In 1946, Purcell demonstrated that spontaneous emission of a quantum emitter is modified when located inside a cavity \cite{Purcell:1946}. A critical parameter is the ratio $Q/V_{eff}$, where Q and $V_{eff}$ refer to the cavity mode quality factor and effective volume, respectively.
In the weak coupling regime, the Purcell factor $F_p$, quantifies the emission rate $\gamma$ inside the cavity compared its free-space value $\gamma_0$ 
\begin{equation}
F_p=\frac{\gamma}{n_1\gamma_0}=\frac{3}{4\pi^2} \left( \frac{\lambda}{n_1} \right )^3 \frac{Q}{V_{eff}} \,,
\label{eq:Purcell}
\end{equation}
where $\lambda$ is the emission wavelength and $n_1$ the cavity optical index. 
When $Q/V_{eff}$ is high enough, strong coupling regime occurs with
reversible energy exchange between the emitter and the cavity mode (Rabi
oscillations) \cite{Vahala:2003}. The design of cavities maximizing this ratio 
in order to control spontaneous emission is extremelly challenging. There is however a
trade-off between Q factor and effective
volume. On one side, ultra high Q ($\sim 10^9$) are obtained 
in microcavities but with large effective volume ($\sim 10^3~\mu m^3$). On the
other side, diffraction limited mode volume [$V_{eff} \sim
(\lambda/n_1)^3$] are achieved in photonic
crystals but at the price of weaker quality factors
($Q\sim10^5$). Moreover, it is sometimes preferable to optimize $Q/V_{eff}$ 
but keeping a reasonable Q factor in order to efficiently extract
the signal from the cavity. Additionally, the emitter spectrum can be large at ambiant temperature 
and better coupling is expected with low Q cavities \cite{Masymov-Hugonin-Lalanne:2010} 
({\it i.-e.} matching cavity and emitter impedances \cite{Greffet-Laroche-Marquier:2010}). 

In this context, it has been proposed to replace the cavity
(polariton) mode by a surface plasmon polariton (SPP) sustained by
metallic structures as an alternative to cavity quantum electrodynamics 
\cite{Chang-Sorensen-Hemmer-Lukin:2006,Cuche-Mollet-Drezet-Huant:2010}. 
SPP can have extremelly reduced effective volume, insuring
high coupling rate with quantum emitters, albeit a poor 
quality factor ($Q\sim 100$ \cite{Cavity}). Particularly, coupling an emitter to a plasmonic wire shed new light on manipulating single photon source at a strongly subwavelength scale, with applications for quantum information processing \cite{qbits}. Others promissing applications deal with the realization of integrated plasmonic amplifier \cite{deLeon-Berini:2008,GrandidierNanolet:2009,ColasdesFrancsOptExp:2010}.  
Highly resolved surface spectroscopy was also pointed out based either on 
the antenna effect \cite{Zuev-Frantsesson-Gao-Eden:2005} or coupling dipolar emission to an optical fiber {\it via} a plasmonic structure \cite{Tanaka-Maletzky-Fischer:2008,Chen-Sandoghdar-Agio:2009}. 
\begin{figure}[h]
\begin{center}
\includegraphics[width=6cm]{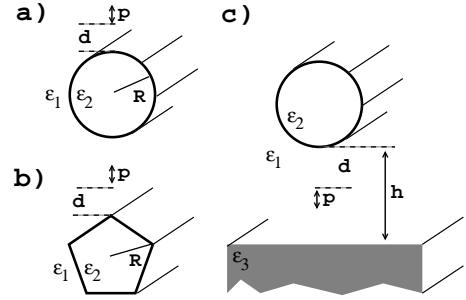}
\caption{\label{schema}Practice models. A dipolar emitter {\bf p} is located at distance $d$ of an infinite silver cylinder of circular (a) or pentagonal (b) cross--section. c) The dipolar emitter is located in a substrate-wire gap.}
\end{center}
\end{figure} 

In this work, we present an original approach for
calculating rigorously the coupling of dipolar emitter to 2D plasmonic 
waveguides of arbitrary profile. We achieve a closed form expression for the
coupling rate into the guided SPP. We also investigate the radiative and non radiative
channels. In particular, the contribution of the plasmon, difficult to estimate otherwise
\cite{Chang-Sorensen-Hemmer-Lukin:2006,Issa-Guckenberger:2007}, is
clearly established. Our method is general and treat equivalently bound and leaky 
waveguides of arbitrary cross-section, possibly on a substrate (Fig.~\ref{schema}). 


According to Fermi's golden rule, coupling of a quantum emitter to a continuum of modes is governed by the (3D) 
local density of states (3D-LDOS)
\begin{equation}
\gamma(\gras r)=\frac{2 \pi \omega}{\hbar \epsilon_0} |p|^2 \rho_{\gras u}(\gras r
,\omega)
\end{equation}
where $\rho_{\gras u}(\gras r ,\omega)$ is the local density
of modes, projected along the direction of the dipolar transition moment $\gras p= p\gras
u$ (partial LDOS) \cite{PLDOS}. $\gras r$ is the emitter location and $\omega$ its emission frequency. 
To characterize the coupling independantly of the emitter properties, we introduce the normalized quantity ${\gamma (\gras r)/\gamma_0}={\rho_{\gras u}(\gras r ,\omega)}/{\rho_{\gras u}^0(\omega)}$ where $\rho_{\gras u}^0(\omega)=\omega^2/6\pi^2c^3$ is the
free-space partial LDOS.

Since we are interested in 2D waveguide, the main idea is to work on
the density of modes associated with the guide (bound and radiation modes). 
For this purpose, we now establish a relationship between 2D and 3D LDOS by introducing Green's dyad formalism.
First, the  3D-LDOS is related to the 3D Green's tensor $\gras G$
of the system ($Im$ and $Tr$ refer to the imaginary part and trace) \cite{PRLChicanneGCF:2001} 
\begin{equation}\label{eqrho}
\rho(\gras r) = -\frac{k_0^2}{\pi \omega} Im Tr \gras G(\gras r,
\gras r) \,.
\end{equation}
In presence of an infinitely long (2D) structure, the 3D-Green's tensor is expressed 
by a Fourier transform
\begin{equation} \label{vide}
\gras G(\gras r,\gras r')=\frac{1}{2\pi}\int_{-\infty}^{\infty} d
k_z \gras G^{2D}(\gras r_\parallel,\gras
r'_\parallel,k_z)e^{-ik_z(z-z')} \,.
\end{equation}
Then, we obtain the 3D-LDOS as a function of 2D-Green's dyad
\begin{equation}
 \rho(\gras r )=-\frac{k_0^2}{2\pi^2 \omega} \int_{-\infty}^\infty dk_z Im Tr\gras G^{2D}
  (\gras r_\parallel,\gras r_\parallel,k_z) \,.
\label{eq:2D23LDOS}
 \end{equation}
Equation (\ref{eq:2D23LDOS}) obviously reproduces the 3D-LDOS in a homogeneous
medium of index $n_1$. Limiting the integration range to radiative waves, 
and since $-\frac{k_0^2}{\pi \omega} Im Tr\gras G^{2D}_{hom}
  (\gras r_\parallel,\gras r_\parallel,k_z)=\omega/2\pi c^2$ in a homogeneous medium, 
we obtain, as expected, $\rho_0(\gras r )=\frac{1}{2\pi } \int_{-n_1k_0}^{n_1k_0} dk_z~
\omega/2\pi c^2 =n_1\omega^2/2\pi^2 c^3$. 
The quantity $-\frac{k_0^2}{\pi \omega}  Im Tr\gras G^{2D}(\gras r_\parallel,\gras r_\parallel,k_z)$ is generally referred as 2D-LDOS by analogy with 3D-LDOS expression (\ref{eqrho}) \cite{2DLDOS}. 
It is a key quantity to understand spatially and spectrally resolved electron energy loss spectroscopy \cite{GarciaAbajo-Kociak:2008a}. 
Equation (\ref{eq:2D23LDOS}) makes then a direct link between 2D and 3D LDOS. We however consider a slightly different definition, more appropriate to describe a density of guided modes \cite{ColasdesFrancsPRB:2009} 
\begin{equation}
\rho^{2D}(\gras r_\parallel,k_z) = -\frac{2k_z}{\pi}Im Tr ~\epsilon(\gras r_\parallel)
\gras G^{2D}  (\gras r_\parallel,\gras r_\parallel,k_z) \,.
\end{equation}
The 2D Green's dyad is separated in two contributions $\gras G^{2D}=\gras G^{2D}_{ref}+\Delta \gras G^{2D}$ where $\gras G^{2D}_{ref}$ is the 2D-Green's dyad without the waveguide and $\Delta \gras G^{2D}$ is the guide contribution. This formulation separates the reference system (multilayer substrate,
homogeneous background, \ldots) from the guiding structure. It comes, with $\epsilon_{ref}$ the dielectric constant of the reference system, 
\begin{eqnarray}
\rho^{2D}(\gras r_\parallel,k_z) 
&=&\rho^{2D}_{ref}(\gras r_\parallel,k_z) + \Delta \rho^{2D}(\gras r_\parallel,k_z)\, , \text{with} \\
\nonumber
\rho^{2D}_{ref}&=&-\frac{2k_z}{\pi}Im Tr ~\epsilon_{ref}(\gras r_\parallel)
\gras G^{2D}_{ref}  (\gras r_\parallel,\gras r_\parallel,k_z) \\
\nonumber \Delta \rho^{2D}&=&-\frac{2k_z}{\pi}Im Tr ~\epsilon(\gras
r_\parallel) \Delta \gras G^{2D}  (\gras r_\parallel,\gras
r_\parallel,k_z).
\end{eqnarray}
This wording separates the continuum of modes of the reference system
$\rho^{2D}_{ref}$ from the waveguide density of modes $\Delta \rho^{2D}$. The partial 2D-LDOS is finally
\begin{equation}
\Delta \rho^{2D}_{\gras u}(\gras r_\parallel,k_z) =
-\frac{2k_z}{\pi}Im Tr~ \epsilon(\gras r_\parallel) [{\gras u} \cdot \Delta \gras G^{2D}(\gras
r_\parallel,\gras r_\parallel,k_z) \cdot {\gras u}]
\label{eq:2DPLDOS}
\end{equation}
\begin{figure}[h!!]
\begin{center}
\includegraphics[width=8cm]{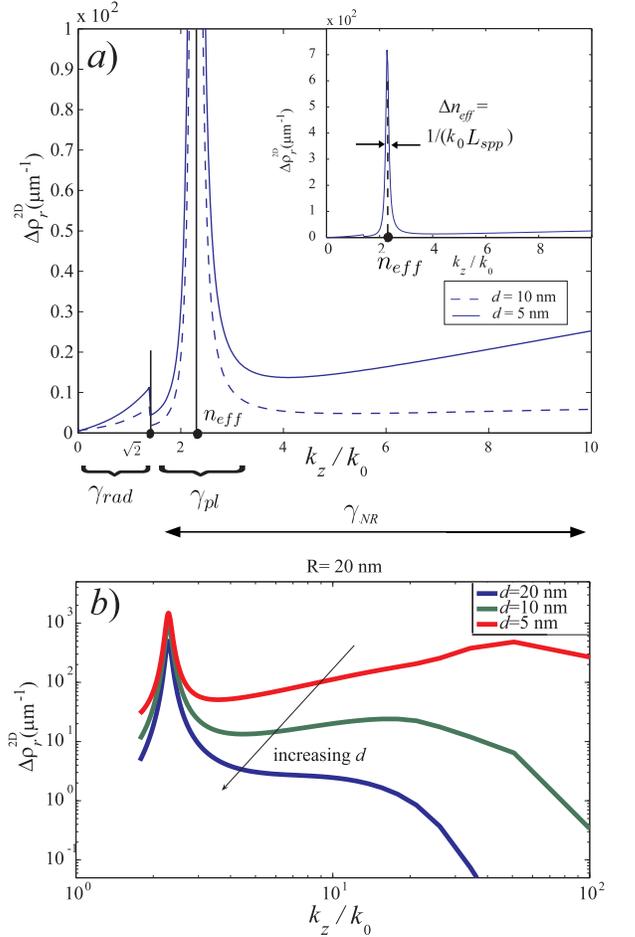}
\caption{\label{LDOS}(Color online) a) 2D radial LDOS variation as a function of $k_z$ at two distances to the nanowire of Fig.~\ref{schema}a). b) Logscale over the high momentum range. $R=20$ nm, $\epsilon_2=-50+3.85i$, $\lambda =1~ \mu m$ and $\epsilon_1=2$.}
\end{center}
\end{figure}
Figure \ref{LDOS} represents the radial 2D-LDOS $\Delta \rho^{2D}_{r}(k_z)$
for the benchmark model defined in Fig.~\ref{schema}a). 2D-Green's dyad has been numerically evaluated by applying a meshing on the waveguide cross-section \cite{ColasdesFrancsPRB:2009}. The main contribution is the Lorentzian variation peaked at the effective index of the guided SPP
$n_{eff}=k_{SPP}/k_0=2.28$, and with a full width at half maximum inversely proportional to the mode propagation length $L_{spp}=1.2$ $\mu$m (inset). 
For $k_z<n_1k_0$, the 2D-LDOS describes scattering events and contributes to radiative rate $\gamma_{rad}$.
Finally, for $k_z>n_1k_0$, LDOS takes part to the non-radiative
decay rate $\gamma_{NR}$. Indeed, the plasmon is dissipated by thermal losses. Moreover, for very short distances, the 2D-LDOS spectrum extends over very large values of $k_z$ (Fig.~\ref{LDOS}b). This behaviour
is typical for non-radiative transfer by electron-hole pairs creation in the metal \cite{Barnes:1998}.

\begin{figure}[h!!]
\begin{center}
\includegraphics[width=8.5cm]{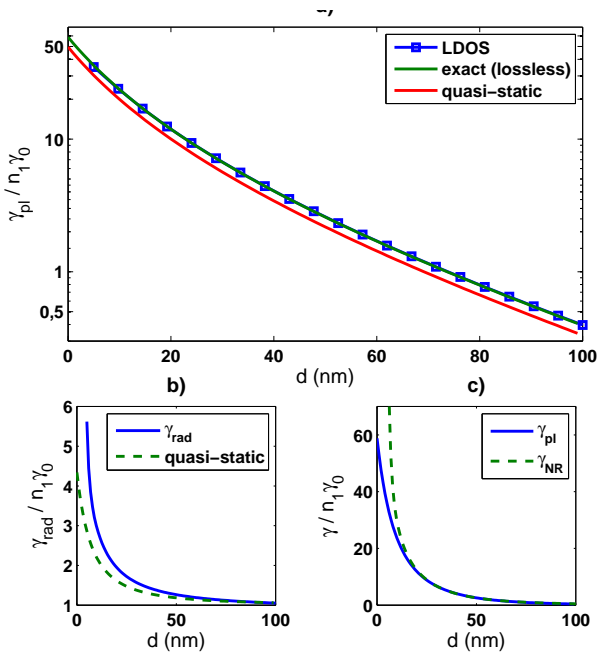}
\caption{\label{fig:GSPP}(Color online) Variation of the rates as a function of distance to the silver nanowire for a radial dipole. a) Coupling rate into SPP obtained using i) our approach based on 2D-LDOS formulation, {\it including losses} ii) exact {\it lossless} case and iii) quasi-static approximation. b) Radiation rate calculated using 2D-LDOS formulation (solid line) or quasi-static approximation (dotted line). c) Comparison of the  plasmon rate $\gamma_{pl}$ with the total non radiative rate $\gamma_{NR}$.}
\end{center}
\end{figure}

The coupling rate into the propagative SPP is obtained using equations (\ref{eqrho},\ref{eq:2D23LDOS},\ref{eq:2DPLDOS}) and keeping only the plasmon contribution 
by limiting the integration of Eq.~(\ref{eq:2D23LDOS}) to $k_z$ corresponding to the SPP resonance. 
This is strongly simplified by the Lorentzian shape of the resonance and leads to the closed form expression \cite{ColasdesFrancsOptExp:2010}
\begin{equation}
\label{eqimportant}
\frac{\gamma_{pl}}{n_1\gamma_0}
=\frac{3\pi \lambda}{4 n_1^3 k_{SPP}} \frac{\Delta \rho^{2D}_{\gras u} (\gras r_\parallel,k_{SPP})}{L_{spp}}  \,.
\end{equation}
This important result describes the emitter coupling rate 
to a 2D waveguide of arbitrary cross section. It is expressed as the overlap between the dipolar emission and the guided mode profile ($\Delta
\rho^{2D}_{\gras u}$) divided by the mode propagation length in the longitudinal direction. 
This defines the 3D Purcell factor for a 2D geometry. Although presented for plasmonic waveguide, the
demonstration remains valid for any 2D configuration
(plasmonic cavity \cite{Cavity}
or waveguide \cite{ColasdesFrancsOptExp:2010}, metal coated
\cite{Masymov-Hugonin-Lalanne:2010} or dielectric \cite{LeKien-Gupta-Hakuta:2005} nanofiber, ...). 
In order to validate this
expression, we now compare it to the exact 
expression obtained by considering coupling to a {\it lossless} waveguide
\cite{LeKien-Gupta-Hakuta:2005,Chen-Nielsen-Gregersen-Lodahl-Mork:2010}:
\begin{equation}
\frac{\gamma_{pl}}{\gamma_0} =\frac{3\pi c E_{\gras u}(d)[E_{\gras
u}(d)]^*}{k_0^2\int_{A_\infty}(\gras{E} \times \gras{H}^*).\gras z.dA}
\label{eq:Lossless}
\end{equation}
where $({\bf E},{\bf H})$ is the electromagnetic field associated with the guided SPP. In Fig.~\ref{fig:GSPP}a), we compare the coupling rate into the plasmonic channel as a function of distance to the silver nanowire obtained using
i) closed form expression (\ref{eqimportant}),
ii) exact expression for a \emph {lossless} plasmonic waveguide (\ref{eq:Lossless}) and
iii) a quasi-static approximation \cite{Chang-Sorensen-Hemmer-Lukin:2006}.

Quite surprinsingly, although the exact expression neglects dissipation, we obtain an excellent agreement with 
our expression that correctly accounts for losses. In formula (\ref{eqimportant}) the ratio $\Delta \rho^{2D}_{\gras u}/L_{spp}$ is proportional to the number of guided modes \cite{ColasdesFrancsPRB:2009} so that it does not depends on the losses. When losses tends towards zero, $L_{SPP}\rightarrow \infty$ and $\Delta \rho^{2D}_{\gras u}\rightarrow \infty$ at resonance so that $\Delta \rho^{2D}_{\gras u}/L_{spp}$ remains constant (Dirac distribution). 
Equivalently, this simply reveals that the emitter couples to the guided mode, no matter if the energy is dissipated by losses during propagation or propagates to infinity.

We now turn on the radiative decay rate associated with the 2D-LDOS in the interval $[-n_1 k_0:n_1 k_0]$. 
We compare in Fig.~\ref{fig:GSPP}b) our numerical simulation with quasi-static approximation derived in Ref.~\cite{Klimov-Lebedev-Ducloy:2004,Chang-Sorensen-Hemmer-Lukin:2006} for the nanowire. The quasi-static approximation underestimates the radiative contribution to the coupling rate since it only considers the cylindrical dipole mode.
\par 
Finally, the non-radiative decay rate $\gamma_{NR}$ is determined from 2D-LDOS calculated on the evanescent domain
$|k_z|> n_1k_0$ which includes all the non radiative mechanisms: Joule losses during plasmon propagation and electron-hole pairs creation into the metal. Figure \ref{fig:GSPP}c) represents the plasmon and total non radiative rates. The non-radiative rate diverges close to the wire surface whereas plasmon contribution remains finite. For large separation distances, the plasmon is the only contribution to the non radiative rate. We achieve an optimal coupling efficiency into the guided SPP, $\beta=\gamma_{pl}/(\gamma_{rad}+\gamma_{NR})=83\%$, at $d=20~nm$. 
\begin{figure}[h!!]
\begin{center}
\includegraphics[width=8.cm]{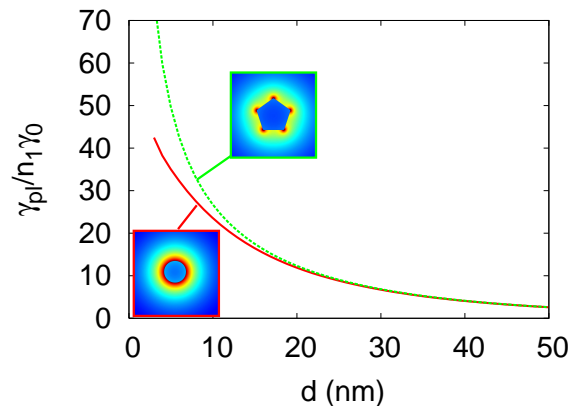}
\caption{\label{fig:pent}Coupling rate to guided SPP calculated near a cylindrical wire of circular (solid line) or pentagonal (dotted line) cross-section ($R=~20 nm$) . The modes profiles are shown.}
\end{center}
\end{figure}
\par
\bigskip
So far, we considered a silver circular nanowire embedded in a homogeneous background to illustrate and validate our method. In the following, we investigate the two complex geometries depicted on Fig.~\ref{schema}b,c). Figure \ref{fig:pent} presents the coupling rate into the SPP supported by a penta-twinned crystalline nanowire recently characterized \cite{Song-Dujardin-Zhang-GCF:2011}. At short distances, the coupling rate into the guided SPP is strongly enhanced as compared to coupling to a circular wire of similar dimensions. This is due to the strong mode confinement near the wire corners as revealed by the mode profile. 
\begin{figure}[h]
\begin{center}
\includegraphics[width=8.cm]{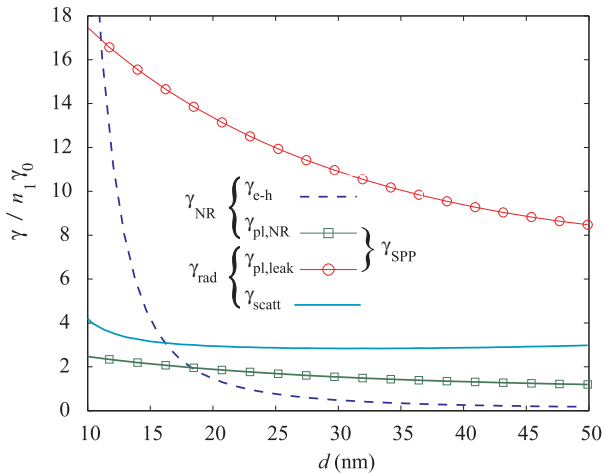}
\caption{\label{fig:leaky}(Color online) Different contributions to the decay rates for a 100 nm diameter silver wire 50 nm above a glass substrate ($\epsilon_3=2.25$). Superstrate is air ($\epsilon_1=1$).}
\end{center}
\end{figure} 
\par
Experimental configurations generally concern structures deposited on a substrate. For high index substrate, the otherwise bound mode becomes leaky. Note that usual expression (\ref{eq:Lossless}) is then practically unenforceable due to difficulty of normalizing the mode. Differently, expression (\ref{eqimportant}), derivated in this work, is easily used even in such a situation. Moreover, in case of leaky mode, it is even more difficult to properly distinguish radiative and non radiative contributions to the coupling rate, as compared to the bound mode situation treated above. Indeed, the guided plasmon contributes to both the radiative rate (leaky part) and non radiative transfer (intrinsic losses). This difficulty is easily overcome using the 2D-LDOS formalism. The propagation length can be written $L_{SPP}=(\Gamma_{rad}^{SPP}+\Gamma_{nrad}^{SPP})^{-1}$ where the radiative and non radiative rates have been introduced.
%
As an example, we consider a 100 nm silver wire 50 nm above a glass substrate. We calculate an effective index $n_{eff}=1.28$, below the substrate optical index, indicating a leaky mode. Its propagation length is $L_{SPP}=1.2~\mu m=1/\Gamma^{SPP}$ with $\Gamma^{SPP}=0.083\mu m^{-1}$. The leakage rate is evaluated by cancelling the metal losses ($Im(\epsilon_2)=0$). We obtain $\Gamma_{rad}^{SPP}=0.073 \mu m^{-1}$. Figure \ref{fig:leaky} shows the interplay between the various contributions to the decay rate for an emitter placed in the wire-substrate gap. The radiative rate $\gamma_{rad}=\gamma_{scatt}+\gamma_{pl,leak}$ is the sum of the scattering and leakage channels and the non radiative rate $\gamma_{NR}=\gamma_{pl,NR}+\gamma_{e-h}$ originates from plasmon losses and electron-hole pairs creation. Except for short distances, the main decay channel is the plasmon decoupling into the substrate. We obtain a maximum decoupling emission into the substrate $\beta=\gamma_{pl,leak}/\gamma=70\%$ for an emitter centered in the gap ($d=25~nm$) \cite{MallekZouari-Buil-Quelin-Mahler-Dubertret-Hermier}.

\par
To conclude, we derive an explicit expression for the coupling rate
between a point--like quantum emitter and a 2D plasmonic waveguide. We
define the coupling Purcell factor into the plasmon channel 
whereas the radiative and non radiative rates are
numerically investigated. This method clearly reveals the physics
underlying the complex mechanisms of spontaneous
emission coupled to a plasmonic guide (scattering, leakage, 
electron-hole pairs creation, SPP excitation). 

This work is supported by French National Agency (ANR PlasTips and $E^2$-Plas). Calculations were performed using DSI-CCUB resources (Universit\'e de Bourgogne).

\end{document}